\documentclass[fleqn,usenatbib]{mnras}
\usepackage[utf8]{inputenc}
\usepackage[T1]{fontenc}
\usepackage{ae,aecompl}
\usepackage{graphicx,amsmath}
\usepackage{color}
\newsavebox{\tablebox}

\newcommand{\kms}{\ifmmode {\rm km\ s}^{-1} \else km s$^{-1}$c\fi}
\newcommand{\lb}{\ifmmode L_{\rm Bol} \else $L_{\rm Bol}$ \fi}
\newcommand{\lxm}{\ifmmode L_{\rm 14-195keV} \else  $L_{\rm 14-195keV}$ \fi}
\newcommand{\mbh}{\ifmmode M_{\rm BH}  \else $M_{\rm BH}$ \fi}
\newcommand{\fx}{\ifmmode F_{\rm X} \else  $F_{\rm X}$ \fi}
\newcommand{\hb}{\ifmmode H\beta \else H$\beta$ \fi}
\newcommand{\ledd}{\ifmmode L_{\rm Edd} \else  $L_{\rm Edd}$ \fi}
\newcommand{\lx}{\ifmmode L_{\rm X} \else  $L_{\rm X}$ \fi}
\newcommand{\ergs}{\ifmmode \rm erg~s^{-1} \else  $\rm erg~s^{-1}$ \fi}
\newcommand{\mdot}{\ifmmode \dot{m} \else \dot{m} \fi }
\newcommand{\msun}{M_{\odot}}

\title[Properties of Hot Corona]{Hot Corona Properties of Swift/BAT detected AGN}
\author[Wang et al.]{Chan Wang, Li-Ming Yu, Wei-Hao Bian\thanks{E-mail: whbian@njnu.edu.cn}, Bi-Xuan Zhao\\
School of Physics and Technology, Nanjing
Normal University, Nanjing 210046, China}

\date{Accepted XXX. Received YYY; in original form \today}

\begin{document}
\maketitle

\begin{abstract}
Using a sample of 208 broad-line active galactic nuclei (AGNs)  from
Swift/BAT AGN Spectroscopic Survey in ultra-hard X-ray band
($14-195$ keV), the hot corona properties are investigated, i.e.
the fraction of gravitational energy dissipated in the hot corona
and the hard X-ray photon index. The bolometric luminosity, \lb,
is calculated from host-corrected luminosity at 5100 \AA. Virial supermassive black hole masses (SMBH, \mbh) are calculated from the  $\rm H\beta$ line width and the corresponding broad line region size-luminosity empirical relation at 5100 \AA. We find a strong anti-correlation between the fraction of energy released in corona ($\fx\equiv \lxm/\lb$) and the Eddington ratio ($ \varepsilon \equiv \lb/\ledd$), $\fx\propto \varepsilon^{-0.60\pm 0.1}$. It is found that this fraction also has a correlation with the SMBH mass,  $\fx\propto \varepsilon^{-0.74\pm 0.14} M_{\rm BH}^{-0.30\pm 0.03}$. Assuming that magnetic buoyancy and feild reconnection lead to the formation of a hot corona, our result favours the shear stress tensor being a proportion of the gas pressure. For our entire sample, it is found that the hard X-ray photon index  $\Gamma$ has a weak but significant correlation with the Eddington ratio, $ \Gamma=2.17+0.21\log \varepsilon$. However, this correlation is not robust because the relation is not statistically significant for its subsample of 32 RM AGNs with relatively reliable \mbh or its subsample of 166 AGNs with single-epoch \mbh. We do not find a statistically significant relation between the photon index and the Eddington ratio taking into account an additional dependence on $\fx$.
\end{abstract}


\section{Introduction}
\label{sec:intro} 
One of the main purposes of studying active
galactic nuclei (AGNs) is to find out how basic features of
supermassive black hole (SMBH) accretion is related to the radiation
field. A model with a hot corona surrounding a cool accretion disk
in AGNs is introduced for the X-ray emission through Compton
up-scattering the disc UV photons by the relative electrons in the
corona \citep{Liang1979, Haardt1991}. There are some kinds of the
corona geometry,  such as the hot planes parallel covering the cold
accretion disk \citep[e.g.][]{Haardt1991, Haardt1994},  a hot sphere
around its central SMBH \citep[e.g.][]{Zdziarski1999}, and an inner
hot sphere plus an inner warm disk \citep{Kubota2018}. A fraction of
total dissipated energy is transferred vertically outside the disk,
and released in the hot, magnetically dominated corona
\citep[e.g.][]{Haardt1991, Svensson1994}. The magnetic field
turbulence have been realized in the transportation of angular
momentum and formation of  the hot corona
\citep[e.g.][]{Merloni2002, Wang2004}.The magnetic stress $t_{\rm
r\phi}$ is assumed be capable of transporting the angular momentum
in the disk. Therefore, the fraction of $\fx \equiv \lx/\lb$ (\lx
and \lb are the X-ray luminosity and the bolometric luminosity
respectively) can be obtained if the magnetic stress and energy
transportation are assumed \citep[e.g.][]{Merloni2002}. This may
give us a possible opportunity to test the working magnetic stress
from hard X-ray observations.

The relation between the fraction of  energy dissipated
in the corona \fx and the Eddington ratio  $\varepsilon$ ($ \varepsilon \equiv
\lb/\ledd$, \ledd is the Eddington luminosity) was investigated by
some authors \citep[e.g.][]{Merloni2002, Wang2004, Yang2006}. For a
compiled sample of 56 AGNs from $ASCA$ observation,
\cite{Wang2004}  found a relation between \fx\ and  $\varepsilon$ as
$\fx\propto \varepsilon^{-0.64\pm0.09}$,  where the X-ray luminosity in
$2-10$ keV is used. Considering a larger sample of 98 AGNs,
\cite{Yang2006}  found $\fx\propto \varepsilon^{-0.66}$. These results
supported the magnetic stress tensor being the form of $t_{\rm
r\phi}\propto P_{\rm gas}$, where $P_{\rm gas}$ is the gas pressure.
In their compiled sample, they only have nine sources with ultra-hard X-ray
observations of INTEGRAL and Swift, and for other sources the luminosity in $2-150$ kev is
extrapolated from the 2-10 keV luminosity with a fixed photon index. 
A sample of more AGNs with direct harder X-ray than 2-10
keV is needed for further investigation, such as Swift/BAT.

Another relation between the X-ray photon index  $\Gamma$ and the
Eddington ratio  $\varepsilon$ was extensively discussed by using different
AGNs samples and by models \citep[e.g.][]{Bian2003,
Wang2004,Yang2006, Brightman13,  Liu2015, Meyer2017, Trakhtenbrot2017}.
\cite{Trakhtenbrot2017}  recently used a sample of 228 hard
X-ray selected low-redshift AGNs drawn from the Swift/BAT AGN
Spectroscopic Survey (BASS) to  investigate this relation. They found a weak but significant correlation
between them, $ \Gamma=log ( \varepsilon ^{0.167})$, and a dependence on the method to derive the SMBH mass.  \cite{Kubota2018}
presented a truncated-disk model including an outer standard cold disk, an inner
warm Comptonising region and a hot
corona  for the broadband spectral energy distribution (SED) of
AGNs (e.g., the soft X-ray excess). They suggested that  $\Gamma$ is also related with \fx, where
increasing of \fx, the curve of $ \Gamma -  \varepsilon$  relation is lower.

In this paper, we use a large sample of 208 low-redshift broad-line AGNs with harder X-ray (14 - 195 kev) emission from Swift/BAT to  further investigate the corona properties. The sample is based on the Swift BASS catalogue which extents harder X-ray at
14-195 keV. This paper is organized as follows. Section 2 presents
our sample. Section 3 is data analysis. Section 4 is our discussion.
Section 5 summaries our results. All of the cosmological
calculations in this paper assume $\Omega_{\Lambda}=0.7$,
$\Omega_{\rm M}=0.3$, and $H_{0}=70~ \kms {\rm Mpc}^{-1}$.

\section{THE SAMPLE}
\label{sec:sample}
\begin{figure*}
\includegraphics[angle=-90,width=3.2in]{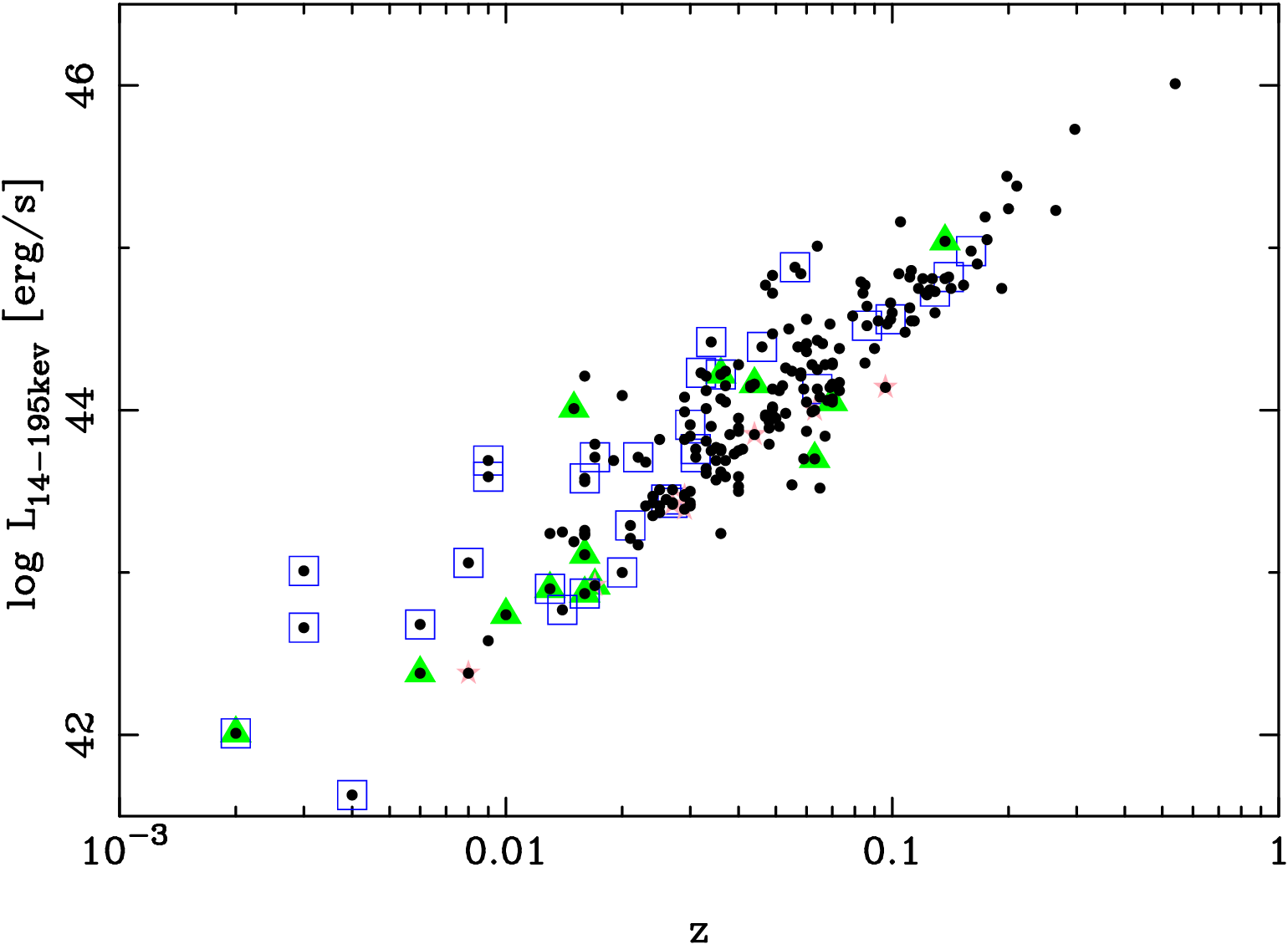}
\includegraphics[angle=-90,width=3.2in]{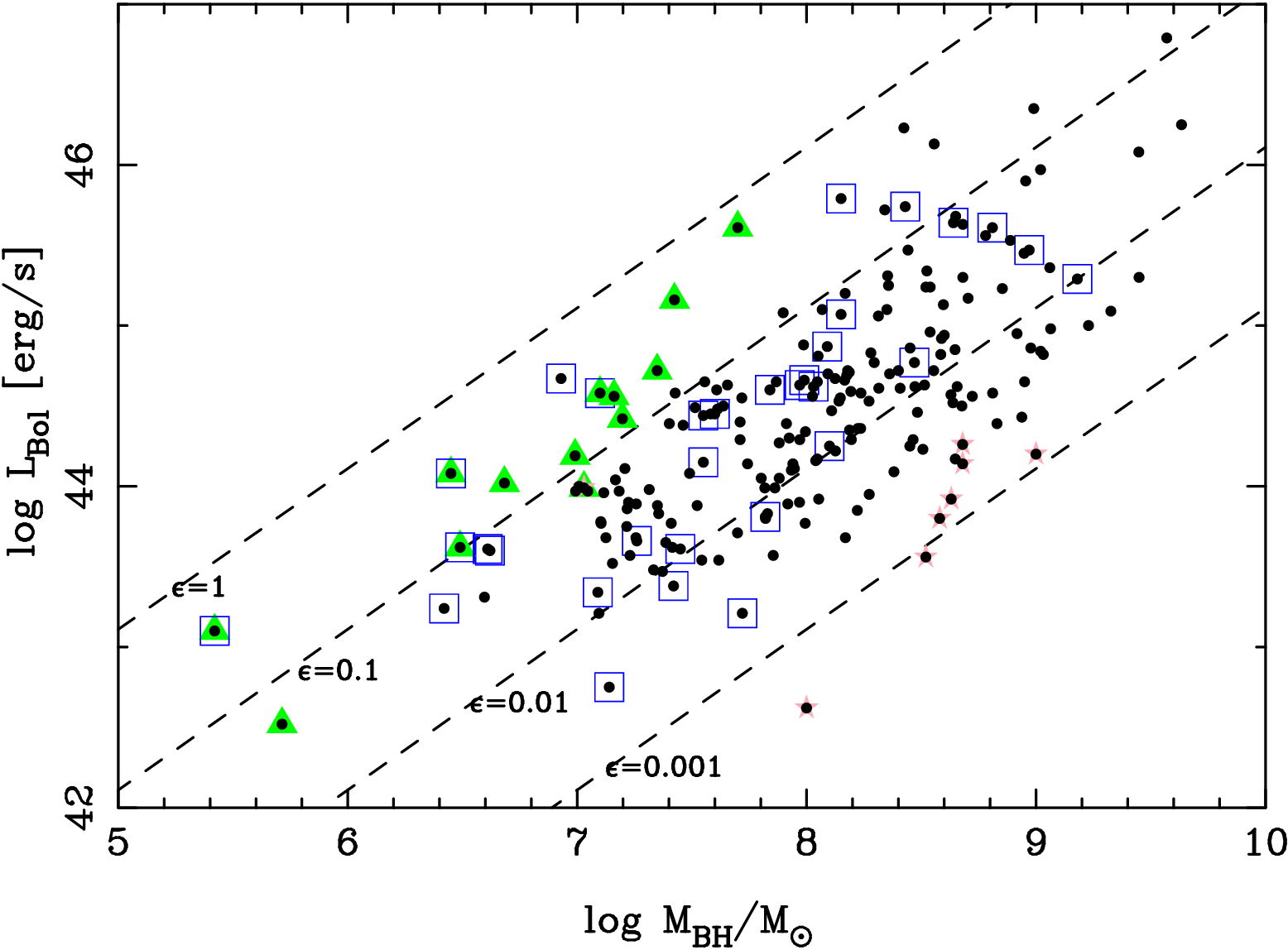}
\caption{Left: The 14-195 keV X-ray luminosity \lxm vs. redshift
$z$. Right: The bolometric luminosity vs. the SMBH mass. The dash
lines show $\varepsilon= \lb/\ledd=1, 0.1, 0.01, 0.001$,
respectively from up to bottom. The black points denote our sample
of 208 broad-line Swift/BAT detected AGNs. Green triangles denote 13
NLS1s. Blue squares represent 34 reverberation mapping source in our
sample. Stars denote 8 AGNs with stellar velocity dispersion.}
\label{Fig1}
\end{figure*}

A sample of AGNs used here is selected from Swift BASS drawn from a
Swift/BAT 70-month catalogue. The Swift/BAT survey
has an all-sky survey in the ultra-hard X-ray band (14-195 kev)
which increases the all-sky sensitivity by a factor of $~20$
compared to previous satellites, such as HEAO 1 \citep{Baumgartner13, Koss2017}.
Most of the Swift/BAT detected AGNs are nearby ($z<0.05$), these
bright and nearby AGNs offer the best opportunity for studies of
corona properties of AGNs with information at ultra-hard X-ray band.

For the Swift BASS, the optical spectroscopic of Swift/BAT sources
(642/836) are from dedicated observations and public archival data
\citep{Koss2017}. According to \cite{Ricci2017Nature}, compared to
the number found at the optical band, the number of broad-line AGNs
decreases significantly at the ultra-hard X-ray band. This is due to
optical central obscuration for these Swift BASS broad-line AGNs.
The X-ray data and the analysis were presented by \cite{Ricci2017}, and we briefly introduce as bellow. For the BASS sample, the analysis by covered the observed-frame energy range of $0.3-150$ keV,  included all the X-ray data available, including Swift/XRT, XMM-Newton/EPIC, Chandra/ACIS, Suzaku/XIS, or ASCA/GIS/SIS observations. The data were modeled with a set of models that rely on an absorbed power-law X-ray SED with a high-energy cut-off, and a reflection component, as well as additional components accounting for warm absorbers, soft excess, Fe $\rm K \alpha$ lines, and/or other spectral features. The typical uncertainty on the hard X-ray photon index  is  less than 0.3 \citep{Ricci2017,Trakhtenbrot2017}. There are 227  Swift/BAT detected broad-line AGNs with measured broad \hb FWHM and the luminosity in 5100 \AA. Excluding 19 beamed sources \citep{Koss2017}, our sample is finally composed of 208 Swift/BAT detected broad-line AGNs. For our sample, the mean value of the uncertainty on the hard X-ray photon index ($\Delta \Gamma$) is 0.15 with a standard deviation of 0.02 \footnote{Two AGNs, i.e. SWIFT J1119.5+5132 and SWIFT J1313.6+3650A, have no $\Gamma$ shown in \cite{Ricci2017} because they were
not observed in the 0.3–10 keV range.}. There are 193 AGNs with $\Delta \Gamma < 0.4$. Considering the number counts larger than 1000 and $0.01<z<0.05$, \cite{Trakhtenbrot2017} presented a sample of 288 AGNs selected from the Swift/BAT to investigate the relation between the hard X-ray photon index and the Eddington ratio. There is a subsample of 126 AGNs with single-epoch spectrum of the \hb broad line in their sample.

The monochromatic luminosity at 5100 \AA\ in the
rest frame, $L_{\rm 5100}$, and the full-width at half-maximum
(FWHM) of H$\beta$, $\rm FWHM_{H\beta}$ are adopted from the Col.
(2) and Col. (4) in Table 9 in \cite{Koss2017}. We present the
properties of our sample of these broad-line AGNs in Table
\ref{table1}. Col. (1) gives the Swift/BAT 70-month hard X-ray
survey ID of the object; Col. (2) is the X-ray luminosity (14-195
keV) in units of erg/s. Col. (1)-(2) are adopted from Table 2 in
\cite{Koss2017}. Col. (6) gives the photon index of the primary
X-ray continuum recovered from the entire energy range (0.3-150 keV) and the full
multi-component model, which is adopted from Col. 3 in Table 5 in
\cite{Ricci2017}.

In the left panel of Fig. \ref{Fig1}, we show \lxm versus $z$ for
our sample. The average value of redshift $z$ is 0.061 with the
standard deviation of 0.057. The average value of $\log \lxm$ is
44.02 with the standard deviation of 0.66 in units of \ergs. In our
sample, there are 32 AGNs with the H$\beta$ lag measured by the
reverberation mapping (RM) method \citep{Du2016}. It is a special
subsample for their reliable SMBH \mbh and host-corrected $L_{\rm
5100}$. There are 8 additional AGNs with measured host velocity
dispersion $\sigma_*$ \citep{Koss2017}. There are 13 narrow-line Seyfert 1 galaxies (NLS1s) with $\rm
FWHM_{\rm H\beta} ~< ~2000 ~km/s$ \citep[e.g.][]{Bian2003}. In Fig. {\ref{Fig1}}, blue
squares, green triangles, stars denote RM AGNs, NLS1s, AGNs with
$\sigma_*$ values, respectively.

\section{ANALYSIS} \label{ANALYSIS}

\begin{figure*}
\includegraphics[angle=-90,width=3.2in]{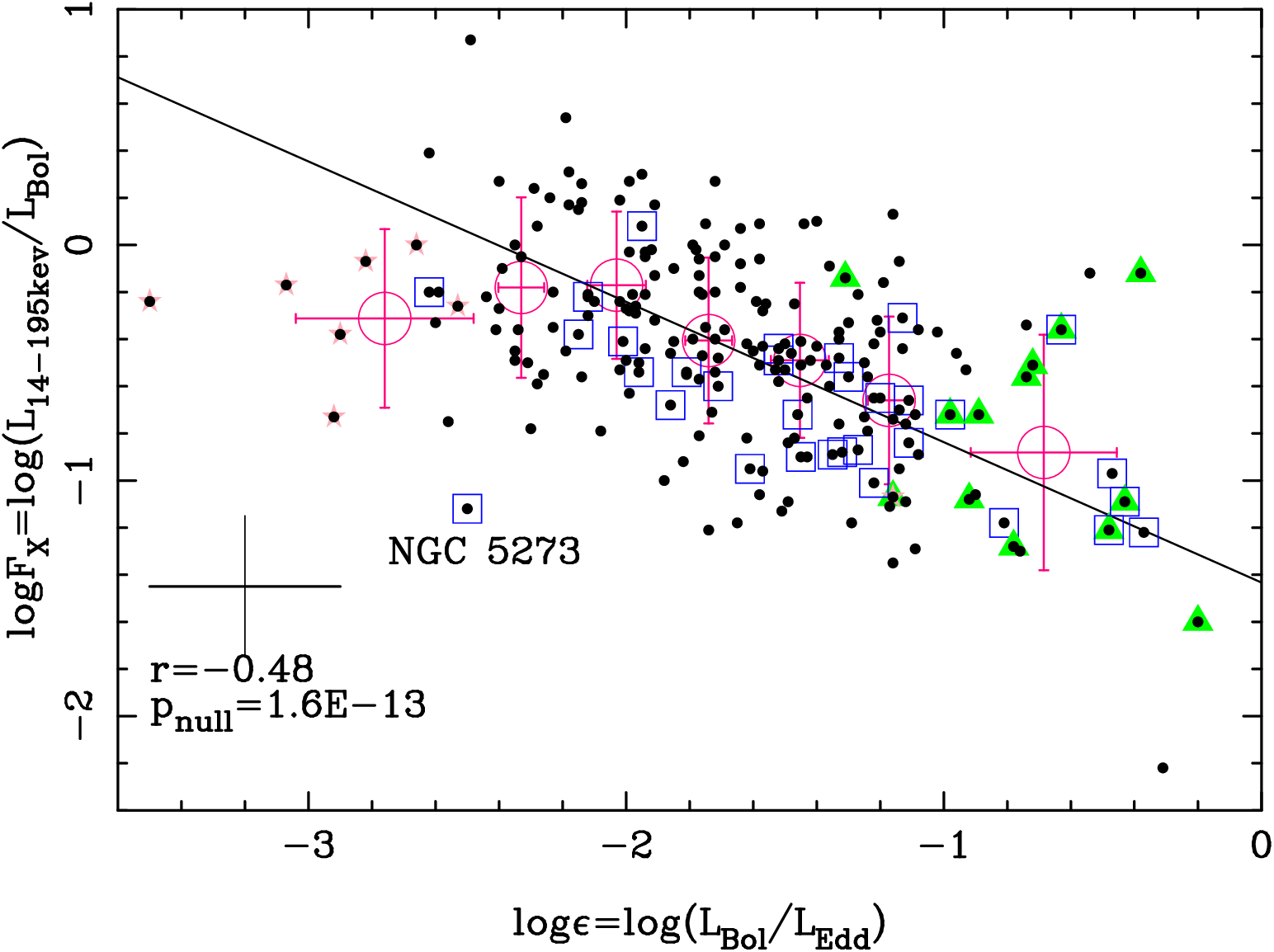}\hfill
\includegraphics[angle=-90,width=3.2in]{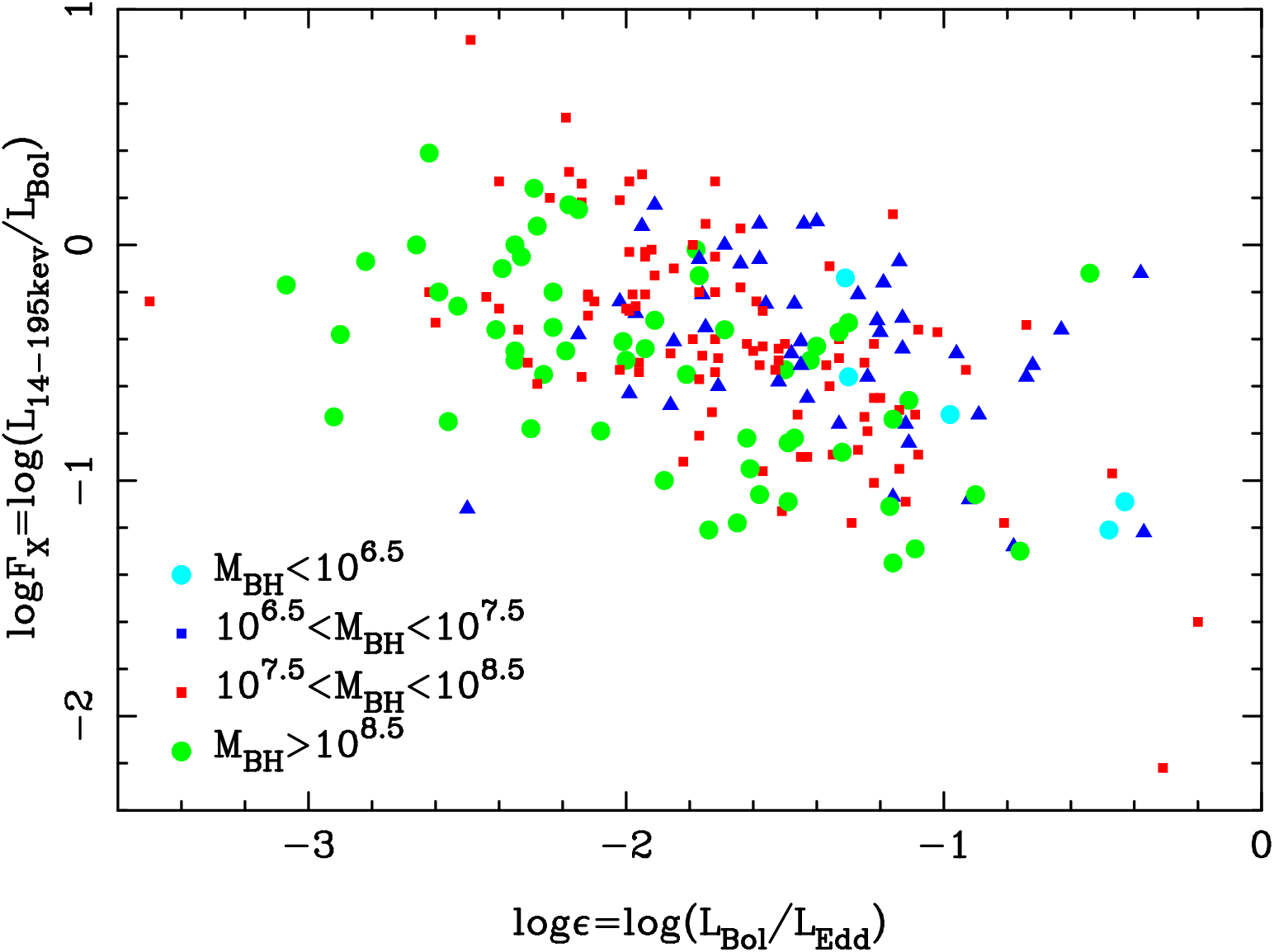}
\caption{Left: \fx vs.  $\varepsilon$. The black points denote our sample of
208 broad-line Swift/BAT detected AGNs. The symbols are the same to
Figure 1. The open red circles denote 7 binned data. The solid black
line is the BCES(Y|X) best-fitting relation. The cross in the lower
left corner represents the typical uncertainties of \fx and  $\varepsilon$. Right:
\fx and  $\varepsilon$ relation with different colors denoting different
ranges of the SMBH mass.} \label{Fig2}
\end{figure*}

\begin{figure}
\centering
\includegraphics[angle=-90,width=3.2in]{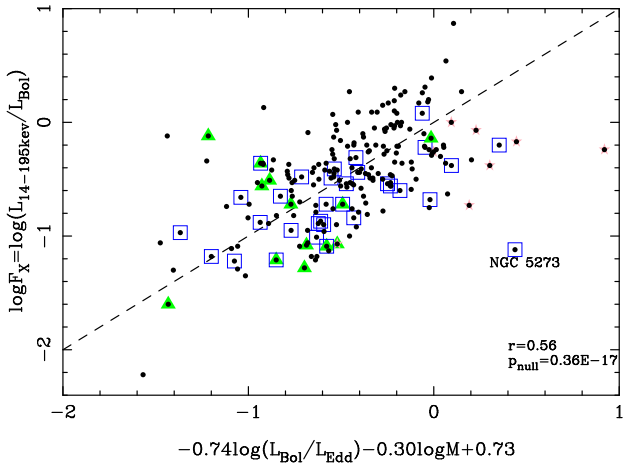}
\caption{Dependence of \fx on the \mbh and the  $\varepsilon$. The symbols
are the same to Figure 1. The dash line is 1:1.} \label{Fig3}
\end{figure}

\subsection{The SMBH mass and the Eddington ratio}
We use this sample of 208 Swift/BAT detected  broad-line AGNs with ultra-hard 14-195 keV
to investigate the relation between the hot corona and the cold
accretion disk. For the hot corona, we use two parameters, i.e. the
fraction of gravitational energy released in the hot corona and the hard X-ray photon index. The SMBH mass and the Eddington ratio are two key
parameters for the SMBH accretion process.

The SMBH masses of broad-line AGNs in our sample are estimated as
follows: (1) for 32 RM AGNs, their RM SMBH masses are preferentially
adopted from \cite{Du2016}; (2) for other 8 AGNs with the stellar
velocity dispersion \citep{Koss2017}, their SMBH masses are
calculated from $M_{\rm BH}-\sigma_{*}$ relation \citep{K13,
Koss2017}; (3) for the rest of 168 AGNs, we calculate their single-epoch
SMBH masses from the empirical $R_{\rm BLR}-L_{\rm 5100}$
relation \citep{Kaspi2000, Bentz2013}. We first remove the host
contribution in $L_{\rm 5100}$ from the empirical formulae by
\cite{Ge2016}. The host fraction $f_{\rm BLR}$ in the total
continuum luminosity at 5100 \AA\ ($L_{\rm 5100}^{total}$) is as
follows,
\begin{equation}
f^{\rm host} = (10.265\pm 2.92)-(0.225\pm0.07)\log L_{\rm
5100}^{total}
\end{equation}
where we do the correction for AGNs with $L_{\rm
5100}^{total} \leq 10^{45.6} \rm erg~ s^{-1}$. The average value of corrected $\log
L_{\rm 5100}$ is 43.51 with the standard deviation of 0.712. The largest correction of
$\log L_{\rm 5100}$ is 0.36 dex. For our subsample of 32 RM AGNs, we directly adopt 
the host-corrected $L_{\rm 5100}$ from the image decomposition \citep{Bentz2013, Du2016}.

Using the H$\beta$ FWHM and the host-corrected $L_{\rm 5100}$, we
calculate the single-epoch SMBH mass
\citep{Kaspi2000,Jun2015,Ge2016},
\begin{equation}
\begin{split}
\mbh & = f \times \rm{FWHM}_{H\rm \beta}^{2}\times  L_{\rm 5100}^{\alpha}/G \\
& =(8.63\pm 2.29) \times 10^{6} \times
L_{5100,44}^{0.533\pm0.034}\rm FWHM_{H\rm \beta,3}^{2} ~M_{\odot}
\end{split}
\end{equation}
where $\rm FWHM_{H\rm \beta,3}$ = $\rm FWHM_{H\rm \beta}/10^{3}\,
kms^{-1}$, $L_{5100,44} = L_{5100}/10^{44}\, \rm erg~s^{-1}$.  In
this formulae, the empirical $R_{\rm BLR}-L_{\rm 5100}$ relation is
adopted from \cite{Bentz2013} where $L_{\rm 5100}$ is the
host-corrected luminosity at 5100\AA, $\alpha=0.533$,
and the factor $f$ is adopted as $f=1.275$ \citep{Woo2013}. The mean
value of $\log (\mbh/M_{\odot})$ in our sample is $8.00$ with the
standard deviation of $0.705$. For the Swift BASS catalogue,
considering the total $L_{\rm 5100}$, the $R_{\rm BLR}-L_{\rm 5100}$
relation with $\alpha=0.65$, $f=1$,  \cite{Koss2017} also calculated
the SMBH masses from the single-epoch spectrum for AGNs with broad \hb lines \citep{Trakhtenbrot2017}. The average value of the mass difference of $\log
\mbh$ between their calculation and ours is 0.1085 dex. The factor
$f$ in our \mbh calculation is adopted as $f=1.275$, the difference of $\log f$
is 0.1055 dex. Therefore, the \mbh difference is mainly from different
adopted factor $f$. The difference is smaller than the \mbh uncertainty of $\sim$ 0.3 dex.

Using the host-corrected luminosity $L_{\rm 5100}$, we calculate the bolometric luminosity \lb through the bolometric correction factor at 5100\AA\ \citep{Marconi2004}. Considering AGN SED from IR to X-ray,
\cite{Marconi2004} gave a formulae about the bolometric correction
factor $f_{\rm bol,b}=\lb/L_{b}$ as a function of the bolometric
luminosity, where $L_{b}$ is the luminosity in B band. We use a
power-law $f_{\nu}\propto\nu^{-0.5}$ converting $L_{b}$ to $L_{\rm
5100}$, and the correction factor formula is,
\begin{equation}
\log(\lb/L_{5100})=0.837-0.067\ell+0.017\ell^{2}-0.0023\ell^{3}
\end{equation}
where $\ell$ = ($\log \lb -12$).  Considering the uncertainties of correction factor $\log(\lb/L_{5100})$ and $\log L_{\rm 5100}$ to be 0.1 dex, the uncertainty of $\log \lb$ is about 0.14 dex \citep{Marconi2004}.  For the range of $L_{\rm 5100}$ in
our sample, $\lb/L_{5100}$ is $\sim$ $18.6 - 6.0$ with a larger
correction factor for a smaller value of $L_{\rm 5100}$. The
Eddington ratio $\varepsilon$ are calculated from the \lb and the
SMBH mass, $ \varepsilon \equiv \lb/\ledd$, mainly in the range from -3 to 0 in $\log~\varepsilon$ scale. The uncertainty of $\log~\varepsilon$ is about 0.33 dex. In Fig. \ref{Fig1}, we show \lb versus \mbh in the right panel for our sample. The dash
lines show $\varepsilon=1, 0.1, 0.01, 0.001$. We has some AGNs with lower $\log~\varepsilon <
-2.5$ comparing with other samples \citep{Wang2004,Yang2006}, where
most of them have stellar velocity dispersion measurements. For 13
NLS1s, they have large Eddington ratios in our sample (Green triangles in Fig. \ref{Fig1}).

\subsection{The relation between the fraction of energy released in
corona and the SMBH accretion}

Using 14-195 keV luminosity by Swift/BAT, we calculate the fraction
of energy dissipated in the corona, i.e. $\fx\equiv \lxm/\lb$. The mean
value of log \fx is -0.445 with the standard deviation of 0.42. It
is slightly larger than the result by \cite{Yang2006}, where 10-150 keV
luminosity are adopted and which was extrapolated from the 2-10 keV luminosity
with a fixed photon index for most of their sources.

The left panel of Fig. \ref{Fig2} shows the \fx versus the Eddington
ratio. We find that \fx has a strong correlation with the Eddington
ratio. The Spearman correlation test gives the Spearman correlation
coefficient $r=-0.48$ and the probability of the null hypothesis
$p_{\rm null} = 1.6\times10^{-13}$. We use the bivariate correlated
errors and scatter method \cite[BCES;][]{AB96}to perform the linear
regression. The BCES(Y|X) best-fitting relation for our total sample
is,
\begin{equation}
\log F_{X}=-(0.60 \pm 0.10) \log \varepsilon - (1.43 \pm 0.17).
\end{equation}
is plotted as solid line in Figure \ref{Fig2}. The uncertainties of $\log~ F_{X}$ and $\log \varepsilon $ are adopted as 0.3 dex, 0.33 dex, respectively. 
Since both \fx and
$\varepsilon$ have a relation with $L_{5100}$, we calculate the correlation
coefficient between \fx and  $\varepsilon$ when $L_{5100}$ is kept fixed. The
partial Kendall correlation coefficient is -0.27, which indicates
that \fx and  $\varepsilon$ are related when excluding the influence of
$L_{5100}$. 32 AGNs with RM SMBH masses are shown as open squares in
the left panel of Figure \ref{Fig2}. The subsample of these 32 RM
AGNs follows this relationship and has a small dispersion compared
with the total sample, except NGC 5273, which has lowest values of $\lxm$  and $L_{\rm 5100}$.
For this subsample, there is a better correlation between \fx and $\varepsilon$ than for the entire sample
with $r=0.53, p_{\rm null}=1.8\times 10^{-3}$. 13 NLS1s with large
Eddington ratios also follow this correlation for the entire sample. If we exclude these 13 NLS1s from the our sample, we find that $r=-0.46,p_{\rm null}=2.0\times 10^{-11}$.
Based on $\varepsilon$, we divide the sample into 7 bins with almost the
same number in each bin. The open circles in the left panel of Fig.
\ref{Fig2} show the mean values of log\fx and $\log \varepsilon$ in each bin;
the error bars show their standard deviations. The BCES(Y|X)
best-fitting relation for binned data is $\log \fx=-(0.354\pm
0.127)\log \varepsilon - (1.04\pm 0.218)$, and the $\fx -  \varepsilon$ relation
curve becomes flat at low Eddington ratio. It is noticed that the
two binned points in the left panel of Fig. \ref{Fig2} are lower
than the entire trend. The flat correlation for binned data is
possibly due to the effect of the SMBH mass just as shown in the
right panel of Fig. \ref{Fig2}.

In the right panel of Fig. \ref{Fig2}, we plot the \fx versus  $\varepsilon$
but with different colors denoting different ranges of the the SMBH
mass. It possibly suggests a selection effect on our sample of Swift/BAT, i.e. higher SMBH mass AGNs can be observed for lower Eddington ratio $\varepsilon$.  Excluding AGNs with smaller $\varepsilon < 0.01$, the relation between \fx and $\varepsilon$ is almost the same with $r=-0.44, p_{\rm null}=1.5\times 10^{-8}$. If additionally excluding 13 NLS1s, we find that $r=-0.42,p_{\rm null}=1.6\times 10^{-7}$. It implies that the selection effect of lower Eddington ratio is not serious in our analysis.  From the right panel of Fig. \ref{Fig2}, it is clear that the relation \fx and  $\varepsilon$ is indeed affected by \mbh. When \mbh increases, \fx decreases. Therefore, we use the
multivariate regression analysis to find the correlation between
\fx,  $\varepsilon$ and \mbh in the form: $\log F_{X}=a+b_{1}\log M_{\rm
BH}+b_{2}\log \varepsilon$. We use the $\chi^{2}$ as the estimator to find
the best values for these fitting parameters
\citep{Merloni2003,Tremaine2002},
\begin{equation}
\chi^{2}=\Sigma_{i}\frac{(y_{i}-a-b_{1}x_{1i}+b_{2}x_{2i})^{2}}{\sigma_{y_{i}}^{2}
+(b_{1}\sigma_{x_{1i}})^{2}+(b_{2}\sigma_{x_{2i}})^{2}}
\end{equation}
,$x_{1},x_{2},y$ correspond to $\log \mbh$, $\log \varepsilon$, $\log \fx$
respectively, $\sigma$ is the corresponding uncertainties. We find the best fit when $\chi^{2}$ is the minimum. About the errors of fitting parameters $
a, b_{1}, b_{2}$, we make the minimum $\chi^{2}/n_{dof}$ unity and then
make $\chi^{2}$-$\chi_{min}^{2}$=1, which corresponds to the error
of 1 $\sigma$. And the fitting result is,
\begin{equation}
 \log \fx=-(0.74^{+0.14}_{-0.14})\log \varepsilon -
(0.30^{+0.03}_{-0.03})\log \mbh + 0.73^{+0.24}_{-0.25} 
\end{equation}
(see Fig. \ref{Fig3}).  The
uncertainties of $\log \mbh$, $\log \varepsilon$, $\log \fx$ are adopted as 0.3 dex, 0.33 dex, 0.3 dex, respectively. In the multivariate regression, $r = 0.56$, $p_{\rm null} =
3.6\times10^{-18}$, which shows that the relationship has been
improved after considering the effect of the SMBH mass. For the
subsample of 32 RM AGNs, it follows this relationship, except NGC
5273 ($r=0.53$, $p_{\rm null}=1.6\times 10^{-3}$). 13 NLS1s with large Eddington ratios also follow this correlation for the
entire sample. This anti-correlation between \fx and  $\varepsilon$, \mbh
indicated that the energy released in corona are decreasing with the
increasing of the Eddington ratio and the black hole mass, 
consistent with Fig. \ref{Fig2}.
Because the Eddington ratio has a dependence on \mbh, the relation among \fx, $\varepsilon$ and \mbh suggests a relation among \fx, \lb and \mbh. Using this $\chi^2$ multivariate regression analysis, we find a relation among \fx, \lb and \mbh, 
\begin{equation}
\log \fx=-(0.67^{+0.01}_{-0.01})\log \lb + (0.57^{+0.04}_{-0.04})\log \mbh  + 24.78^{+0.34}_{-0.34} 
\end{equation}
The uncertainties of $\log \mbh$, $\log \lb$, $\log \fx$ are adopted as 0.3 dex, 0.14 dex, 0.3 dex, respectively. In this multivariate regression, $r = 0.53$, $p_{\rm null} =
1.1\times10^{-16}$.

\subsection{The relation between the photon index and the Eddington ratio}

\begin{figure}
\includegraphics[angle=-90,width=3.2in]{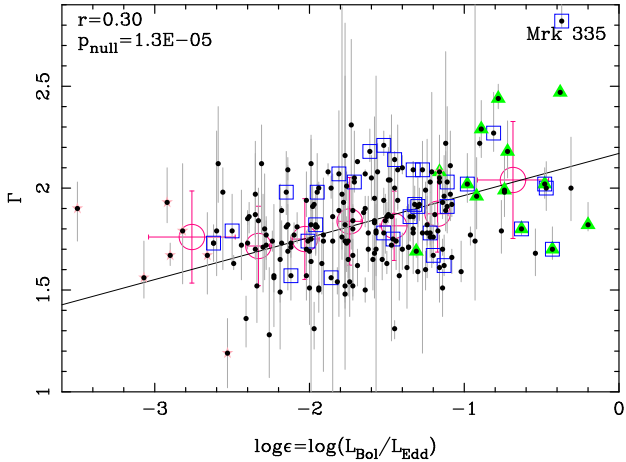}
\caption{
$\Gamma$ vs.  $\varepsilon$. The solid black line is the BCES(Y|X) best-fitting relation. The symbols are the same to Fig. \ref{Fig1}.
 } \label{Fig4}
\end{figure}

Fig. \ref{Fig4} shows the hard X-ray photon index $\Gamma$ versus the Eddington ratio  $\varepsilon$ for our Swift/BAT broad-line AGNs. The Spearman correlation coefficient between them is $r=0.30$
and $p_{\rm null} = 1.3\times10^{-5}$. This correlation  is statistically significant, albeit weak. The best linear fitting of
BCES(Y|X) is,
\begin{equation}
\Gamma = (0.21\pm 0.05)\log \varepsilon + (2.17\pm 0.09)
\end{equation}
The uncertainty of $\log \varepsilon $ is adopted as 0.33 dex. It is consistent with the result by \cite{Trakhtenbrot2017}, where
the \lb was estimated from hard X-ray luminosity instead.  13 NLS1s with large
Eddington ratios also follow this correlation for the entire sample. If we exclude these 13 NLS1s from our total sample, it makes the parameter range of $\varepsilon$ smaller, and we find that the correlation becomes weaker with $r=0.24, p_{\rm null}=6.3\times 10^{-4}$. If excluding AGNs with $\Delta \Gamma > 0.4$ from our total sample, there are 193 AGNs and the correlation becomes slightly stronger with $r=0.31, p_{\rm null}=1.2\times 10^{-5}$. Only for 166 AGNs with single-epoch \mbh, $r=0.25, p_{\rm null}=1.5\times 10^{-3}$, the correlation is no too significant. Considering the possible effects of spectroscopic aperture \citep{Trakhtenbrot2017}, for our subsample of 80 AGNs at with single-epoch \mbh at $0.05<z<0.5$, $r=0.33, p_{\rm null}=2.7\times 10^{-3}$. Excluding AGNs with smaller $\varepsilon < 0.01$ from the possible selection effect, this correlation becomes weaker with $r=0.22, p_{\rm null}=6.2\times 10^{-3}$. It is due to smaller parameter range of $\varepsilon$, just like that for NLS1s. 
Thereofore, for the subsample of AGNs with single-epoch \mbh, the correlation is not too significant, which is consistent with the result by \cite{Trakhtenbrot2017} (their $p_{\rm null}=1.1\times 10^{-3}$). For the subsample of 32 RM AGNs, this correlation is weaker with $r=0.25, p_{\rm null}=0.165$. It is consistent with the result by \cite{Trakhtenbrot2017}. Mrk 335 is an outline from these 32 RM AGNs, which has a ultra-soft X-ray spectrum (i.e. highest $\Gamma$) in our sample. 

Based on $\varepsilon$, we also divide the sample into 7 bins with almost the same
number in each bin. The open red circles in Fig.
\ref{Fig4} show the mean values of $\Gamma$ and log $\varepsilon$ in each
bin; the error bars show their standard deviations. It is found that
the binned data with the smallest  $\varepsilon$ seems deviate from this
relationship.

For our sample, we also find that $\Gamma$ has no significant correlation with $\lb$ ($r=0.11, p_{\rm null}=0.11$), or with \mbh ($r=-0.11, p_{\rm null}=0.11$) , which is consistent with the result with \cite{Trakhtenbrot2017}.

Considering uncertainties of $\Gamma$, $\varepsilon$ and \fx, we also do the $\chi^2$ multivariate regression analysis. However, because of large uncertainties on these parameters, we can not find a suitable multivariate regression with a large $p_{\rm null}$ ( $r=0.07, p_{\rm null}=0.3$).


\section{DISCUSSION}
\subsection{Energy released in the corona \fx and the magnetic stresses tensor}
It is assumed that angular momentum transport was carried out by turbulence and that the stress tensor $t_{r\phi}$ was scaled to the disk pressure $P$, $t_{r\phi}=\alpha P$, $\alpha$ is the viscosity parameter \citep{SS73}. The viscous stress was suggested to be generally proportional to the magnetic pressure through numerical simulations. It is believed that the strong buoyancy and magnetic field reconnection inevitably lead to the formation of a hot corona \citep{Stella1984}.
Through magnetic buoyancy, the fraction of gravitational energy
dissipated in the hot corona was calculated for different accretion mode \citep[e.g.][]{Merloni2002, Wang2004, Yang2006}. 
The relationship between \fx and  $\varepsilon$ can
be used to test the working stress of magnetic field turbulence
\citep[e.g.][]{Stella1984, Merloni2002, Wang2004}. 
The fraction of the energy transported by
magnetic buoyancy to be $\fx=P_{mag}v_{p}/Q$, where $v_{p}$ is the
transporting velocity, $P_{mag} = B^{2}/8\pi$ is the magnetic
pressure, and the dissipated energy $Q=-(3/2)c_{\rm s}t_{\rm r\phi}$
\citep{Merloni2002}. 
The viscous stress is assumed to scale with magnetic pressure, $t_{r\phi}=-k_0 P_{mag}$, we have 
\begin{equation}
\fx=\frac {2v_{p}}{3k_{0}c_{s}} =
\frac{2^{\frac{3}{2}}b}{3k_{0}}\sqrt{\frac{P_{mag}}{P_{tot}}} =
\frac{2^{\frac{3}{2}}b}{3k_{0}^{\frac{3}{2}}}\sqrt{\frac{(-t_{r\phi})}{P_{tot}}}
= C\sqrt{\frac{(-t_{r\phi})}{P_{tot}}}
\end{equation}
where $b = v_{p}/v_{A}$, $ Alfv\acute{e}n$ velocity $v_{A} =
B/\sqrt{4\pi\rho}$, $P_{tot} = P_{rad} + P_{gas}$, $c_{s}^{2}
=P_{tot}/\rho$ and $C = 2^{3/2}b/(3k_{0}^{3/2})\sim 1$. Therefore,
if the magnetic stress is assumed, we can get the $\fx$ at every
radius and for different accretion rate. A global value of $<\fx>$ can
then be obtained by integrating over all the disc area for different
magnetic stress tensor \citep{Svensson1994, Merloni2002, Wang2004,
Yang2006}. \cite{Yang2006} calculated theoretical light curves for
the relation between the factor $f$ and the Eddington ratio for six
distinct types of $t_{r \phi}$, which has relations with $P_{gas}, P_{rad}, P_{total}$ \citep{Wang2004,Yang2006}.
The stress tensors of $t_{r\phi}$ are $-\alpha P_{total}$, $-\alpha P_{gas}$, $-\alpha P_{rad}$, $-\alpha \sqrt{P_{rad} P_{total}}$, $-\alpha \sqrt{P_{gas} P_{total}}$, $-\alpha \sqrt{P_{gas} P_{rad}}$ for the models from 1 to 6, respectively. $\alpha$ is the viscosity. \cite{Wang2004} adopted $\alpha=0.05, 0.1, 0.4$ for $\mbh=10^8 \msun$, and $\alpha = 0.05$ but for $\mbh=10^{6} \msun$. \cite{Yang2006} adopted $\alpha=0.1, 0.8$ for $\mbh=10^8, 10^6 \msun$. Comparing with the calculation by \cite{Wang2004}, \cite{Yang2006} considered advection cooling and thermal instability in their calculation. The advection cooling has an effect on $\fx$ at dimensionless accretion rate $\mdot >1$ \citep[see right panel in Fig. 5 in][]{Yang2006}. Their results are consistent with each other.  Different models have different slope of the $\fx-\varepsilon$ relation. This relation is moved along the y-axis of $\fx$ for different \mbh and $\alpha$. However, the slope of this relation is not sensitive to them. For model 1, the slope is zero. For model 3 or 4, the slope is positive. For model 2 or 5 the slope is negative. For model 6, the slope changes from a positive value to a negative value \citep[see Fig. 5 in][]{Yang2006}. For all the models, $\fx$ is proportional to $\alpha^{0.5}$. Considering our large value of $\fx$, large $\alpha$ is needed ($\alpha=0.8$). We use the slope of the relation between $\fx$ and $\varepsilon$ to distinguish different models.
For their model 2, i.e. magnetic stress tensor $t_{r \phi} \propto P_{gas}$, $\fx \propto \varepsilon^{-0.77}$. For their model 5, i.e. magnetic stress tensor
$t_{r \phi} \propto \sqrt{P_{gas}P_{total}}$, $\fx \propto \varepsilon^{-0.44}$ \citep[see also][]{Wang2004}. For our sample, $\fx \propto  \varepsilon^{-0.60\pm 0.1}$. It is slightly flatter than $-0.64, -0.66$ by \cite{Wang2004,Yang2006},
respectively for AGNs samples with only 2-10 keV data.
However, considering the effect of \mbh, $\fx \propto  \varepsilon^{-0.74\pm 0.14}$. The steeper index of $-0.72$  favours model 2 in \cite{Yang2006}, where the magnetic stress tensor is $t_{r\phi} \propto {P_{gas}}$.

Comparing with result by \cite{Wang2004}, as shown in Fig.
\ref{Fig2}, we find that the $\fx-\varepsilon$ relation depends on
the SMBH masses, $\fx \propto \mbh^{-0.30\pm 0.03}$. Differences in
\mbh may flatten the slope as shown as the binned data in the left
panel of Fig. \ref{Fig2}. The trend of \mbh dependence on the
$\fx-\varepsilon$ relation is also consistent with the theoretical
curves calculated assuming the same viscosity of $\alpha=0.1$ by \cite{Yang2006} (see their Fig. 5). For AGNs with small SMBH masses, it is expected that \fx should be large for these AGNs with the small Eddington ratio \lb/\ledd.

\subsection{Hard X-ray Photon index $\Gamma$}
We present the relation between the hard X-ray photon index  $\Gamma$ and
the Eddington ratio  $\varepsilon$ for our sample of 208 low-reshift
broad-line AGNs from Swift BASS catalogue. We find a correlation
between  $\Gamma$ and $ \varepsilon$: $ \Gamma=(2.17\pm 0.09)+(0.21\pm 0.05) \log  \varepsilon$. The
relation index of $0.21\pm 0.05$ is consistent with $0.26\pm 0.05$,
$0.26\pm 0.05$, $0.17\pm 0.04$ by
\cite{Bian2003,Wang2004,Trakhtenbrot2017}, respectively. Although we find this correlation is significant for our entire sample, the correlation for a subsample of 166 AGNs with sing-epoch \mbh or a subsample of 32 RM AGNs with relatively reliable \mbh is not significant. Therefore, this relation between $\Gamma$ and $\varepsilon$ is statistically significant but not robust, which depends on the \mbh determination method, and more reliable \mbh is needed in the future for this correlation analysis.
\cite{Yang2015} found a V-type relation between  $\Gamma$ and  $\varepsilon$, and
the turning point is at $\log \varepsilon\simeq-3$, which is due to the
change of the accretion mode. In Fig. \ref{Fig4}
we find a possible turning point at $\log \varepsilon\simeq-3$ for our
binned data. 

According to the study of \cite{Kubota2018}, they
develop a truncated-disk model for the broad-band SED of AGNs which
includes an outer standard disc, an inner warm Comptonising region
and a hot corona. The inner warm Comptonising region has different
parameters from the corona, in electron temperature ($ kT_e\sim 0.1-1
~ \rm{keV} ~vs.~ 40-100~\rm{keV}$) and the optical depth ($\rm \tau \sim 10-25
~vs.~ 1-2$) \citep{Kubota2018}. In their theoretical calculation for this truncated-disk geometry,
they found that  $\Gamma$ - $\dot m$ relation is affected by \fx, with
the increasing of \fx, the relation curve of $ \Gamma - \dot m$
relation is lower (see their Fig. 5), where $\dot m$ is the
dimensionless accretion rate (i.e. \lb/\ledd assuming a constant
efficient). 
By employing a multivariate $\chi^2$ regression technique, taking uncertainties into account, we investigate the possible presence of a relation amongst $\varepsilon$, \fx and $\Gamma$ as predicted by the model of a warm Comptonising corona proposed by \cite{Kubota2018}. We do not find a statistically significant relation amongst these parameters, which is, at face value, in contradiction with the prediction by \cite{Kubota2018}. Yet, we caution that the absence for such a relation in our dataset can be partly due to the large uncertainties on these parameters.


\section{CONCLUSIONS}
Using a compiled sample of 208 broad-line AGNs from Swift/BAT
Spectroscopic Survey with ultra-hard X-ray band (14-195kev)
observations, the properties of hot corona are investigated. The
main conclusions can be summarized as follows:
\begin{itemize}
\item For our sample of low-redshift broad-line AGNs, host-corrected $L_{\rm 5100}$ is used to
estimate \lb, and empirical $R_{\rm BLR}-L_{\rm 5100}$ relation by
\cite{Bentz2013} is used to calculate the single-epoch SMBH \mbh,
except for AGNs with measured RM \mbh and host stellar velocity
dispersion. There is a subsample of 32 RM AGNs with reliable \mbh
and host-corrected $L_{\rm 5100}$, and a subsample of 13 NLS1s. The
fraction of gravitational energy dissipated in the hot corona is
estimated from 14-195 keV by Swift, $\fx \equiv \lxm/\lb$. For our
sample , the mean value of log \fx is -0.445 with the standard
deviation of 0.42.

\item It is found that \fx is both correlated with the Eddington ratio and the black hole mass,
$\fx\propto \varepsilon^{-0.74\pm 0.14}\mbh^{-0.30\pm 0.03}$, which indicates that the energy released in corona are decreasing with the increasing of the Eddington ratio and the black hole mass. The subsample of 32 RM AGNs also follows this correlation, as well as for 13 NLS1s. Considering magnetic buoyancy and magnetic field reconnection leading to the formation of a hot corona, our result favors the magnetic stress tensor being a proportion of the gas pressure, $t_{r\phi} \propto {P_{gas}}$, which is consistent with the result by \cite{Wang2004}.

\item  For our total sample, the hard X-ray photon index has a correlation with the Eddington ratio  $\varepsilon$, $ \Gamma=(2.17\pm 0.09)+(0.21\pm 0.05)\log \varepsilon$. This correlation is statistically significant, albeit weak. However, this correlation is not robust because the relation for its subsample of 32 RM AGNs with relatively reliable \mbh or its subsample of 166 AGNs with single-epoch \mbh is not significant. Therefore, the significant of this relation between $\Gamma$ and $\varepsilon$ depends on the \mbh determination method and more reliable \mbh is needed in the future for this correlation analysis. Considering large uncertainties of $\varepsilon$, $\fx$ and $\Gamma$, from the $\chi^2$ multivariate regression analysis, we do not find a statistically significant relation between the photon index and the Eddington ratio taking into account an additional dependence on $\fx$.


\end{itemize}


\section*{Acknowledgements}
We are very grateful to Wang Jian-Min for the instructive comments. We are also very grateful to the anonymous referee for her/his instructive comments which significantly improved the content of the paper. This work is supported by the National Key Research and Development Program of China (No. 2017YFA0402703). This work has been supported by the National Science Foundations of China (Nos. 11373024, 11233003, and 11873032).

\newpage
\begin{table*}

\begin{center}
\caption[]{A sample of 208 broad-line AGNs detected by Swift
BAT.\\
Col.(1): Swift/BAT 70-month hard X-ray survey ID; Col.(2-3): The
luminosity of hard X-ray in 14-195 keV and host-corrected $5100$
\AA\ in units of erg/s; Col.(4): The bolometric luminosity \lb in
units of erg/s; Col.(5): Black hole mass; Col.(6): The photon index
adopted from \citep{Ricci2017}; Col.(7): Notes, D: $L_{\rm 5100}$
from \cite{Du2016}; K: $L_{\rm 5100}$ from \cite{Koss2017}; 1: \mbh
from \cite{Du2016}; 2: \mbh from the $M_{\rm BH}-\sigma_*$ relation
with measured $\sigma_*$ \citep{K13,
Koss2017}; 0: \mbh from the single-epoch spectrum in this work. \\
}
\label{table1}
\begin{lrbox}{\tablebox}
\begin{tabular}{ccccccc}
\hline
ID & $\log L_{\rm x(14-195kev)}$ & $\log L_{\rm 5100~\AA}$ & $\log \lb$ &$\log \mbh$ &  $\Gamma$ & Notes \\
   & $\rm erg~s^{-1}$ & $\rm erg~s^{-1}$ & $\rm erg~s^{-1}$ & $\rm M_\odot$ &  & \\
 (1)&(2)&(3)&(4)&(5)&(6)&(7)\\
\hline
     6&43.45&43.76&44.67&6.93&$ 2.82^{+ 0.08}_{- 0.03}$&D 1\\
   16&44.82&44.97&45.79&8.15&$ 2.00^{+ 0.13}_{- 0.11}$&D 1\\
   36&44.41&43.13&44.11&7.95&$ 1.81^{+ 0.36}_{- 0.08}$&K 0\\
   39&44.98&44.81&45.64&8.64&$ 2.03^{+ 0.37}_{- 0.06}$&D 1\\
   43&43.19&42.40&43.48&7.33&$ 1.92^{+ 0.03}_{- 0.02}$&K 0\\
   60&43.97&42.87&43.88&7.35&$ 1.68^{+ 0.50}_{- 0.22}$&K 0\\
   61&44.26&42.99&43.99&7.86&$ 1.82^{+ 0.28}_{- 0.06}$&K 0\\
   73&44.39&43.98&44.87&8.09&$ 2.09^{+ 0.04}_{- 0.03}$&D 1\\
   77&42.92&42.99&43.99&7.03&$ 2.08^{+ 0.18}_{- 0.08}$&K 2\\
   94&43.41&42.74&43.77&7.99&$ 1.52^{+ 0.16}_{- 0.18}$&K 0\\
   99&43.93&43.71&44.63&7.65&$ 1.93^{+ 0.13}_{- 0.08}$&K 0\\
  106&44.15&43.62&44.55&8.15&$ 1.84^{+ 0.08}_{- 0.06}$&K 0\\
  111&44.60&44.21&45.09&9.33&$ 1.97^{+ 0.26}_{- 0.21}$&K 0\\
  113&44.47&43.80&44.71&8.19&$ 1.65^{+ 0.12}_{- 0.15}$&K 0\\
  116&43.43&43.50&44.44&7.55&$ 1.78^{+ 0.08}_{- 0.06}$&D 1\\
  127&44.05&43.63&44.56&8.03&$ 1.83^{+ 0.26}_{- 0.14}$&K 0\\
  129&43.56&42.74&43.77&7.41&$ 2.01^{+ 0.07}_{- 0.07}$&K 0\\
  130&42.87&43.10&44.08&6.45&$ 2.02^{+ 0.08}_{- 0.07}$&D 1\\
  134&44.14&43.19&44.17&8.05&$ 2.12^{+ 0.14}_{- 0.23}$&K 0\\
  147&44.72&42.83&43.85&8.22&$ 1.63^{+ 0.04}_{- 0.03}$&K 0\\
  162&43.64&43.44&44.39&8.83&$ 1.98^{+ 0.12}_{- 0.38}$&K 0\\
  167&44.41&43.25&44.22&8.13&$ 1.94^{+ 0.14}_{- 0.08}$&K 0\\
  169&43.99&43.61&44.53&8.14&$ 1.80^{+ 0.20}_{- 0.16}$&K 0\\
  183&44.24&43.75&44.66&8.17&$ 1.70^{+ 0.26}_{- 0.11}$&K 0\\
  190&43.59&43.35&44.30&7.92&$ 2.31^{+ 0.42}_{- 0.61}$&K 0\\
  197&43.57&42.98&43.98&7.31&$ 1.31^{+ 0.04}_{- 0.05}$&K 0\\
  213&43.89&43.34&44.29&7.97&$ 1.97^{+ 0.92}_{- 0.27}$&K 0\\
  214&44.83&44.38&45.25&8.36&$ 1.76^{+ 0.22}_{- 0.10}$&K 0\\
  220&44.84&44.34&45.20&8.17&$ 1.92^{+ 0.05}_{- 0.06}$&K 0\\
  223&43.52&43.73&44.65&8.05&$ 1.84^{+ 0.13}_{- 0.16}$&K 0\\
  224&44.16&44.07&44.95&8.92&$ 1.81^{+ 0.07}_{- 0.08}$&K 0\\
  228&43.69&43.55&44.48&7.61&$ 1.78^{+ 0.46}_{- 0.22}$&K 0\\
  230&43.75&43.29&44.25&8.45&$ 2.12^{+ 0.19}_{- 0.34}$&K 0\\
  232&43.85&43.40&44.35&8.19&$ 1.81^{+ 0.17}_{- 0.13}$&K 0\\
  242&43.23&42.39&43.47&7.37&$ 1.70^{+ 0.09}_{- 0.08}$&K 0\\
  244&44.08&43.41&44.36&8.24&$ 1.75^{+ 0.09}_{- 0.07}$&K 0\\
  254&44.21&44.23&45.10&8.07&$ 1.66^{+ 0.04}_{- 0.08}$&K 0\\
  261&43.79&42.56&43.62&7.42&$ 1.77^{+ 0.04}_{- 0.02}$&K 0\\
  266&44.23&43.87&44.77&8.47&$ 2.07^{+ 0.04}_{- 0.04}$&D 1\\
  269&43.24&42.48&43.54&7.54&$ 1.73^{+ 0.02}_{- 0.01}$&K 0\\
  274&44.02&44.36&45.23&8.85&$ 1.54^{+ 0.08}_{- 0.17}$&K 0\\
  285&43.35&42.64&43.68&8.17&$ 1.79^{+ 0.49}_{- 0.22}$&K 0\\
  291&43.21&42.97&43.97&7.18&$ 1.80^{+ 0.16}_{- 0.08}$&K 0\\
  301&42.58&42.09&43.21&7.09&$ 1.64^{+ 0.60}_{- 0.65}$&K 0\\
  310&44.09&43.16&44.14&7.74&$ 1.89^{+ 0.02}_{- 0.02}$&K 0\\
  313&43.90&42.97&43.97&6.99&$ 1.70^{+ 0.05}_{- 0.03}$&K 0\\
  314&45.04&44.29&45.16&7.42&$ 2.47^{+ 0.02}_{- 0.02}$&K 0\\
  316&44.21&43.44&44.39&7.91&$ 1.50^{+ 0.06}_{- 0.06}$&K 0\\
  318&44.38&43.65&44.58&8.24&$ 1.82^{+ 0.73}_{- 0.36}$&K 0\\
  338&44.55&44.18&45.06&8.31&$ 1.61^{+ 0.30}_{- 0.20}$&K 0\\
  347&43.69&43.69&44.61&8.32&$ 1.54^{+ 0.07}_{- 0.06}$&K 0\\
  363&43.87&43.93&44.83&8.28&$ 1.70^{+ 0.85}_{- 1.02}$&K 0\\
  376&44.75&44.62&45.47&8.44&$ 1.97^{+ 0.11}_{- 0.13}$&K 0\\
  378&43.50&43.45&44.40&7.71&$ 2.09^{+ 0.22}_{- 0.19}$&K 0\\
  382&43.71&43.68&44.60&7.84&$ 1.86^{+ 0.08}_{- 0.12}$&D 1\\
  384&43.43&42.64&43.68&7.12&$ 2.08^{+ 0.14}_{- 0.26}$&K 0\\
  389&43.47&43.23&44.20&9.00&$ 1.93^{+ 0.17}_{- 0.10}$&K 2\\

\hline
\end{tabular}

\end{lrbox}
\scalebox{0.9}{\usebox{\tablebox}}
\end{center}
\end{table*}

\newpage
\begin{table*}

\begin{center}
\setcounter{table}{0}
\caption[]{--continu}
\begin{lrbox}{\tablebox}
\begin{tabular}{ccccccc}
\hline
ID & $\log L_{\rm x(14-195kev)}$ & $\log L_{\rm 5100~\AA}$ & $\log \lb$ &$\log \mbh$ &  $\Gamma$ & Notes \\
   & $\rm erg~s^{-1}$ & $\rm erg~s^{-1}$ & $\rm erg~s^{-1}$ & $\rm M_\odot$ &  & \\
 (1)&(2)&(3)&(4)&(5)&(6)&(7)\\
\hline

  394&44.75&44.44&45.30&9.45&$ 1.28^{+ 0.33}_{- 0.21}$&K 0\\
  398&44.28&43.81&44.72&8.55&$ 1.51^{+ 0.03}_{- 0.03}$&K 0\\
  403&43.96&42.99&43.99&7.82&$ 1.50^{+ 0.25}_{- 0.11}$&K 0\\
  409&44.56&44.91&45.74&8.43&$ 2.27^{+ 0.20}_{- 0.09}$&D 1\\
  411&43.75&42.95&43.96&7.12&$ 1.78^{+ 0.23}_{- 0.59}$&K 0\\
  418&44.74&43.57&44.50&8.68&$ 1.80^{+ 0.27}_{- 0.25}$&K 0\\
  420&44.55&44.11&45.00&9.23&$ 1.87^{+ 0.30}_{- 0.10}$&K 0\\
  425&45.44&44.45&45.31&8.35&$ 2.03^{+ 0.08}_{- 0.10}$&K 0\\
  431&44.81&44.51&45.36&9.06&$ 1.89^{+ 0.49}_{- 0.52}$&K 0\\
  443&43.69&42.72&43.75&7.22&$ 1.84^{+ 0.07}_{- 0.07}$&K 0\\
  447&44.39&44.85&45.68&8.65&$ 2.11^{+ 0.17}_{- 0.20}$&K 0\\
  449&43.82&43.92&44.82&8.59&$ 2.08^{+ 0.04}_{- 0.02}$&K 0\\
  455&44.01&43.43&44.38&7.46&$ 1.76^{+ 0.06}_{- 0.03}$&K 0\\
  458&44.22&43.66&44.58&7.10&$ 1.80^{+ 0.04}_{- 0.03}$&D 1\\
  459&43.53&43.38&44.34&7.99&$ 2.16^{+ 0.65}_{- 1.05}$&K 0\\
  460&43.48&43.05&44.04&7.17&$ 1.82^{+ 0.14}_{- 0.10}$&K 0\\
  461&43.42&42.77&43.80&8.58&$ 1.67^{+ 0.06}_{- 0.05}$&K 2\\
  466&44.72&43.91&44.81&8.05&$ 1.71^{+ 0.21}_{- 0.09}$&K 0\\
  470&43.17&42.45&43.52&7.15&$ 1.65^{+ 0.38}_{- 0.18}$&K 0\\
  473&44.64&43.63&44.56&8.72&$ 1.77^{+ 0.09}_{- 0.12}$&K 0\\
  481&43.24&43.00&44.00&7.01&$ 1.89^{+ 0.29}_{- 0.10}$&K 0\\
  495&43.70&43.34&44.29&8.46&$ 1.72^{+ 0.12}_{- 0.12}$&K 0\\
  497&42.66&42.24&43.34&7.09&$ 1.56^{+ 0.01}_{- 0.01}$&D 1\\
  501&44.77&44.44&45.30&8.68&$ 1.86^{+ 0.07}_{- 0.04}$&K 0\\
  507&44.36&43.70&44.62&8.47&$ 1.31^{+ 0.13}_{- 0.10}$&K 0\\
  512&43.89&43.18&44.16&8.04&$ 1.69^{+ 0.08}_{- 0.07}$&K 0\\
  524&43.62&42.81&43.83&7.83&$ 1.71^{+ 0.06}_{- 0.05}$&K 0\\
  529&44.75&44.37&45.24&8.54&$ 1.57^{+ 0.19}_{- 0.12}$&K 0\\
  530&43.59&42.79&43.81&7.82&$ 1.57^{+ 0.13}_{- 0.12}$&D 1\\
  532&43.39&42.50&43.56&8.52&$ 1.56^{+ 0.18}_{- 0.10}$&K 2\\
  537&44.48&43.69&44.61&8.41&$\cdots$                 &K 0\\
 542&43.29&42.55&43.60&6.62&$ 1.62^{+ 0.15}_{- 0.05}$&D 1\\
  549&43.85&42.92&43.92&8.63&$ 1.79^{+ 0.33}_{- 0.26}$&K 2\\
  550&46.01&45.32&46.13&8.56&$ 1.68^{+ 0.12}_{- 0.11}$&K 0\\
  552&43.43&42.88&43.89&7.26&$ 2.04^{+ 0.22}_{- 0.17}$&K 0\\
  556&44.28&43.79&44.70&8.09&$ 1.63^{+ 0.06}_{- 0.05}$&K 0\\
  558&43.69&42.56&43.61&7.45&$ 1.98^{+ 0.04}_{- 0.04}$&D 1\\
  561&44.25&43.73&44.65&7.87&$ 1.86^{+ 0.04}_{- 0.08}$&K 0\\
  565&43.73&42.94&43.95&8.27&$ 1.72^{+ 0.27}_{- 0.17}$&K 0\\
  566&42.38&41.25&42.52&5.71&$ 1.69^{+ 0.03}_{- 0.03}$&K 0\\
  567&44.12&43.65&44.58&7.43&$ 1.77^{+ 0.04}_{- 0.05}$&K 0\\
  572&44.14&43.71&44.63&8.51&$ 1.51^{+ 0.31}_{- 0.18}$&K 0\\
  575&44.14&43.16&44.14&8.68&$ 1.67^{+ 0.21}_{- 0.19}$&K 2\\
  576&43.95&43.51&44.45&7.58&$ 1.99^{+ 0.04}_{- 0.11}$&K 0\\
  583&43.00&42.29&43.38&7.42&$ 1.98^{+ 0.20}_{- 0.11}$&D 1\\
  585&42.01&41.96&43.10&5.42&$ 1.70^{+ 0.11}_{- 0.05}$&D 1\\
  587&44.90&44.03&44.92&8.59&$ 1.93^{+ 0.05}_{- 0.10}$&K 0\\
  589&44.00&43.30&44.26&8.68&$ 1.19^{+ 0.17}_{- 0.17}$&K 2\\
  595&43.01&42.09&43.21&7.72&$ 1.73^{+ 0.03}_{- 0.03}$&D 1\\
  596&43.41&42.87&43.88&7.52&$ 1.73^{+ 0.13}_{- 0.14}$&K 0\\
  607&42.38&41.37&42.62&8.00&$ 1.90^{+ 0.13}_{- 0.16}$&K 2\\
  608&42.90&42.57&43.62&6.49&$ 2.02^{+ 0.15}_{- 0.09}$&D 1\\
  611&44.28&44.80&45.63&8.68&$ 2.05^{+ 0.10}_{- 0.09}$&K 0\\
  613&43.47&42.67&43.71&7.70&$ 1.77^{+ 0.06}_{- 0.04}$&K 0\\
  623&44.13&43.70&44.62&8.03&$ 2.21^{+ 0.07}_{- 0.05}$&D 1\\
  624&44.60&44.08&44.96&8.54&$ 1.62^{+ 0.08}_{- 0.10}$&K 0\\
  631&43.06&42.62&43.66&7.26&$ 2.03^{+ 0.01}_{- 0.01}$&D 1\\
  636&43.82&43.11&44.09&8.38&$ 1.73^{+ 0.16}_{- 0.03}$&K 0\\
  641&42.77&42.56&43.61&6.61&$ 1.91^{+ 0.18}_{- 0.13}$&D 1\\
  644&44.13&43.79&44.70&8.36&$ 1.48^{+ 0.06}_{- 0.07}$&K 0\\
  651&44.28&43.73&44.65&7.56&$ 1.59^{+ 0.15}_{- 0.05}$&K 0\\
  652&43.84&42.51&43.57&7.86&$ 1.85^{+ 0.03}_{- 0.03}$&K 0\\
  657&43.26&42.74&43.77&7.10&$ 2.00^{+ 0.10}_{- 0.12}$&K 0\\
\hline
\end{tabular}

\end{lrbox}
\scalebox{0.9}{\usebox{\tablebox}}
\end{center}
\end{table*}

\newpage
\begin{table*}

\begin{center}
\setcounter{table}{0}
\caption[]{--continu}
\begin{lrbox}{\tablebox}
\begin{tabular}{ccccccc}
\hline
ID & $\log L_{\rm x(14-195kev)}$ & $\log L_{\rm 5100~\AA}$ & $\log \lb$ &$\log \mbh$ &  $\Gamma$ & Notes \\
   & $\rm erg~s^{-1}$ & $\rm erg~s^{-1}$ & $\rm erg~s^{-1}$ & $\rm M_\odot$ &  & \\
 (1)&(2)&(3)&(4)&(5)&(6)&(7)\\
\hline

  663&43.50&43.10&44.08&7.49&$ 1.94^{+ 0.06}_{- 0.07}$&K 0\\
  664&43.84&43.06&44.05&7.88&$\cdots$                 &K 0\\
  667&44.29&43.73&44.65&8.95&$ 1.36^{+ 0.11}_{- 0.14}$&K 0\\
  683&43.95&43.44&44.39&7.40&$ 1.96^{+ 0.20}_{- 0.07}$&K 0\\
  686&41.63&41.54&42.75&7.14&$ 1.79^{+ 0.18}_{- 0.16}$&D 1\\
  689&44.29&43.81&44.72&8.18&$ 1.78^{+ 0.18}_{- 0.21}$&K 0\\
  694&44.21&42.89&43.90&7.97&$ 1.89^{+ 0.02}_{- 0.01}$&K 0\\
  695&43.61&42.85&43.86&7.22&$ 1.72^{+ 0.15}_{- 0.12}$&K 0\\
  697&43.91&43.71&44.63&7.97&$ 1.75^{+ 0.11}_{- 0.05}$&D 1\\
  702&44.12&42.92&43.92&8.05&$ 1.74^{+ 0.05}_{- 0.05}$&K 0\\
  713&44.66&43.96&44.86&8.98&$ 1.57^{+ 0.17}_{- 0.08}$&K 0\\
  717&43.71&43.29&44.25&8.10&$ 1.82^{+ 0.10}_{- 0.15}$&D 1\\
  719&43.68&42.64&43.68&7.26&$ 2.13^{+ 0.01}_{- 0.01}$&K 0\\
  722&44.38&43.96&44.86&8.45&$ 2.05^{+ 0.23}_{- 0.07}$&K 0\\
  726&44.63&44.60&45.45&8.95&$ 1.78^{+ 0.12}_{- 0.10}$&K 0\\
  728&44.52&44.63&45.47&8.97&$ 2.18^{+ 0.11}_{- 0.05}$&D 1\\
  734&43.59&43.06&44.05&7.80&$ 2.00^{+ 0.06}_{- 0.05}$&K 0\\
  735&43.76&43.74&44.66&7.99&$ 2.14^{+ 0.04}_{- 0.03}$&D 1\\
  741&43.98&42.77&43.80&7.82&$ 1.82^{+ 0.29}_{- 0.25}$&K 0\\
  744&44.71&44.69&45.53&8.89&$ 1.86^{+ 0.50}_{- 0.61}$&K 0\\
  748&44.13&43.27&44.23&8.51&$ 1.86^{+ 0.07}_{- 0.08}$&K 0\\
  750&43.11&43.22&44.19&6.99&$ 1.96^{+ 0.19}_{- 0.16}$&K 0\\
  753&44.05&43.12&44.10&7.93&$ 2.00^{+ 0.03}_{- 0.02}$&K 0\\
  754&43.76&43.34&44.29&7.71&$ 1.65^{+ 0.10}_{- 0.10}$&K 0\\
  760&44.16&43.81&44.72&7.35&$ 1.99^{+ 0.18}_{- 0.09}$&K 0\\
  765&44.53&44.10&44.98&9.06&$ 1.65^{+ 0.37}_{- 0.21}$&K 0\\
  769&44.81&44.06&44.94&8.60&$ 1.74^{+ 0.28}_{- 0.13}$&K 0\\
  774&43.71&43.17&44.15&7.55&$ 1.78^{+ 0.07}_{- 0.12}$&D 1\\
  776&44.53&43.95&44.85&8.65&$ 1.74^{+ 0.31}_{- 0.64}$&K 0\\
  779&45.23&45.16&45.97&9.02&$ 1.61^{+ 0.16}_{- 0.17}$&K 0\\
  793&44.74&43.64&44.57&8.63&$ 1.73^{+ 0.11}_{- 0.08}$&K 0\\
  795&43.24&42.60&43.65&7.39&$ 1.74^{+ 0.09}_{- 0.20}$&K 0\\
  797&44.73&44.77&45.61&8.81&$ 1.92^{+ 0.09}_{- 0.08}$&D 1\\
  801&43.77&43.57&44.50&7.64&$ 1.60^{+ 0.09}_{- 0.10}$&K 0\\
  846&44.50&44.23&45.10&8.35&$ 1.71^{+ 0.12}_{- 0.05}$&K 0\\
  862&44.12&43.16&44.14&7.94&$ 1.63^{+ 0.12}_{- 0.15}$&K 0\\
  882&45.38&44.89&45.72&8.34&$ 1.98^{+ 0.12}_{- 0.12}$&K 0\\
  883&43.87&42.75&43.78&7.10&$ 1.74^{+ 0.11}_{- 0.14}$&K 0\\
  905&44.15&42.88&43.89&7.92&$ 2.09^{+ 0.06}_{- 0.05}$&K 0\\
  912&44.82&43.49&44.43&8.94&$ 1.73^{+ 0.09}_{- 0.09}$&K 0\\
  923&43.97&43.19&44.17&8.65&$ 2.12^{+ 0.28}_{- 0.19}$&K 0\\
  924&43.54&43.56&44.49&7.51&$ 1.51^{+ 0.09}_{- 0.14}$&K 0\\
  925&43.70&43.48&44.42&7.20&$ 2.29^{+ 0.14}_{- 0.14}$&K 0\\
  948&44.07&43.31&44.27&7.88&$ 1.52^{+ 0.12}_{- 0.07}$&K 0\\
  950&44.58&43.65&44.58&8.81&$ 1.70^{+ 0.11}_{- 0.05}$&K 0\\
  967&45.73&46.01&46.79&9.57&$ 2.22^{+ 0.03}_{- 0.02}$&K 0\\
  984&44.84&43.86&44.77&8.29&$ 2.07^{+ 0.02}_{- 0.01}$&K 0\\
  994&44.88&44.43&45.29&9.18&$ 1.74^{+ 0.06}_{- 0.03}$&D 1\\
  995&43.25&42.89&43.90&7.22&$ 1.94^{+ 0.05}_{- 0.05}$&K 0\\
 1013&44.72&43.81&44.72&8.40&$ 1.76^{+ 0.12}_{- 0.11}$&K 0\\
 1032&44.24&43.78&44.69&8.18&$ 2.05^{+ 0.11}_{- 0.11}$&K 0\\
 1036&44.06&43.94&44.84&9.02&$ 1.71^{+ 0.26}_{- 0.22}$&K 0\\
 1040&44.43&43.61&44.53&8.27&$ 1.86^{+ 0.21}_{- 0.95}$&K 0\\
 1041&44.23&43.99&44.88&7.99&$ 2.03^{+ 0.05}_{- 0.05}$&K 0\\
 1042&42.74&43.03&44.02&6.68&$ 2.44^{+ 0.07}_{- 0.05}$&K 0\\
 1043&43.75&42.81&43.83&7.35&$ 1.90^{+ 0.06}_{- 0.09}$&K 0\\
 1045&44.15&43.41&44.36&8.22&$ 1.91^{+ 0.10}_{- 0.08}$&K 0\\
 1046&42.68&42.12&43.24&6.42&$ 1.91^{+ 0.14}_{- 0.10}$&D 1\\
 1063&43.41&42.21&43.31&6.60&$ 1.70^{+ 0.08}_{- 0.10}$&K 0\\
 1084&43.79&43.13&44.11&7.21&$ 2.08^{+ 0.39}_{- 0.47}$&K 0\\
 1088&45.16&45.45&46.25&9.63&$ 2.04^{+ 0.16}_{- 0.21}$&K 0\\
 1089&44.01&45.43&46.23&8.42&$ 2.00^{+ 0.25}_{- 0.16}$&K 0\\
 1090&44.42&44.19&45.07&8.15&$ 1.67^{+ 0.09}_{- 0.04}$&D 1\\

\hline
\end{tabular}

\end{lrbox}
\scalebox{0.9}{\usebox{\tablebox}}
\end{center}
\end{table*}

\newpage
\begin{table*}

\begin{center}
\setcounter{table}{0}
\caption[]{--continu}
\begin{lrbox}{\tablebox}
\begin{tabular}{ccccccc}
\hline
ID & $\log L_{\rm x(14-195kev)}$ & $\log L_{\rm 5100~\AA}$ & $\log \lb$ &$\log \mbh$ &  $\Gamma$ & Notes \\
   & $\rm erg~s^{-1}$ & $\rm erg~s^{-1}$ & $\rm erg~s^{-1}$ & $\rm M_\odot$ &  & \\
 (1)&(2)&(3)&(4)&(5)&(6)&(7)\\
\hline

1099&43.51&43.68&44.60&7.61&$ 2.22^{+ 0.45}_{- 0.33}$&K 0\\
 1102&44.79&45.08&45.90&8.96&$ 1.79^{+ 0.04}_{- 0.04}$&K 0\\
 1104&44.77&43.92&44.82&9.03&$ 1.84^{+ 0.16}_{- 0.14}$&K 0\\
 1106&43.90&43.52&44.46&8.48&$ 1.69^{+ 0.23}_{- 0.19}$&K 0\\
 1110&44.01&44.78&45.61&7.70&$ 1.82^{+ 0.11}_{- 0.03}$&K 0\\
 1117&43.51&42.51&43.57&7.23&$ 1.52^{+ 0.16}_{- 0.08}$&K 0\\
 1118&44.39&43.76&44.67&8.12&$ 1.79^{+ 0.13}_{- 0.06}$&K 0\\
 1120&45.24&45.27&46.08&9.45&$ 1.87^{+ 0.03}_{- 0.03}$&K 0\\
 1122&44.55&44.20&45.08&7.90&$ 1.74^{+ 1.34}_{- 0.29}$&K 0\\
 1132&45.05&45.55&46.35&8.99&$ 1.79^{+ 0.15}_{- 0.11}$&K 0\\
 1146&44.07&44.26&45.13&8.60&$ 1.95^{+ 0.40}_{- 0.17}$&K 0\\
 1151&44.56&43.54&44.47&8.11&$ 1.74^{+ 0.17}_{- 0.18}$&K 0\\
 1153&44.86&43.66&44.59&8.19&$ 1.67^{+ 0.08}_{- 0.20}$&K 0\\
 1162&43.37&43.62&44.55&7.72&$ 1.92^{+ 0.08}_{- 0.08}$&K 0\\
 1168&44.08&42.48&43.54&7.62&$ 1.49^{+ 0.51}_{- 0.28}$&K 0\\
 1172&45.01&44.49&45.34&8.52&$ 1.59^{+ 0.35}_{- 0.05}$&K 0\\
 1176&44.05&43.63&44.56&7.16&$ 2.18^{+ 0.15}_{- 0.13}$&K 0\\
 1178&43.81&42.97&43.97&7.05&$ 1.87^{+ 0.06}_{- 0.06}$&K 0\\
 1182&43.58&43.51&44.45&7.60&$ 2.09^{+ 0.14}_{- 0.09}$&D 1\\
 1183&44.77&43.70&44.62&8.66&$ 1.81^{+ 0.01}_{- 0.01}$&K 0\\
 1185&43.76&43.34&44.29&8.19&$ 1.57^{+ 0.13}_{- 0.12}$&K 0\\
 1187&44.17&43.59&44.52&8.64&$ 1.56^{+ 0.36}_{- 0.24}$&K 0\\
 1189&43.99&44.31&45.17&8.70&$ 1.88^{+ 0.03}_{- 0.03}$&K 0\\
 1195&44.81&44.37&45.24&8.52&$ 2.10^{+ 0.06}_{- 0.03}$&K 0\\
 1206&45.19&44.72&45.56&8.78&$ 1.78^{+ 0.35}_{- 0.20}$&K 0\\
\hline
\end{tabular}

\end{lrbox}
\scalebox{0.9}{\usebox{\tablebox}}
\end{center}
\end{table*}

\end{document}